# Defect energetics of concentrated solid-solution alloys from *ab initio* calculations: $Ni_{0.5}Co_{0.5}$, $Ni_{0.5}Fe_{0.5}$, $Ni_{0.8}Fe_{0.2}$ and $Ni_{0.8}Cr_{0.2}$

Shijun Zhao,[1] G. Malcolm Stocks,[1] and Yanwen Zhang[1,2]

[1] Materials Science and Technology Division, Oak Ridge National Laboratory, Oak Ridge, Tennessee 37831, USA
[2] Department of Materials Science and Engineering, University of Tennessee, Knoxville, Tennessee 37996, USA

## Abstract

It has been shown that concentrated solid solution alloys possess unusual electronic, magnetic, transport, mechanical and radiation-resistant properties that are directly related to underlying chemical complexity. Because every atom experiences a different local atomic environment, the formation and migration energies of vacancies and interstitials in these alloys exhibit a distribution, rather than a single value as in a pure metal or dilute alloy. Using *ab initio* calculations based on density functional theory and special quasirandom structure, we have characterized the distribution of defect formation energy and migration barrier in four Ni-based solid-solution alloys: $Ni_{0.5}Co_{0.5}$, $Ni_{0.5}Fe_{0.5}$, $Ni_{0.8}Fe_{0.2}$, and $Ni_{0.8}Cr_{0.2}$. As defect formation energies in finite-size models depend sensitively on the elemental chemical potential, we have developed a computationally efficient method for determining it which takes into account the global composition and the local short-range order. In addition we have compared the results of our *ab initio* calculations to those obtained from available embedded atom method (EAM) potentials. Our results indicate that the defect formation and migration energies are closely related to the specific atomic size in the structure, which further determines the elemental diffusion properties. Different EAM potentials yield different features of defect energetics in concentrated alloys, pointing to the need for additional potential development efforts in order to allow spatial and temporal scale-up of defect and simulations, beyond those accessible to *ab initio* methods.

# 1 Introduction

Recently, high entropy alloys (HEAs) comprising four, five or more metallic elements at or near an equiatomic ratio that form single-phase solid-solutions on simple underlying face-centered-cubic (*fcc*) or body-centered-cubic (*bcc*) lattices have been reported.[1] They are distinctly different from conventional alloys, which are typically designed based on one or two principal elements. Previous investigations have shown that HEAs possess many unusual properties, such as good thermal stability,[2] high mechanical performance,[3] and good fatigue and corrosion resistance,[4, 5] as well as exceptional strength and ductility at cryogenic temperatures.[6] Even more recently, it was found that a series of Ni containing single-phase binary, ternary and quaternary concentrated solid-solution alloys also exhibit extraordinary properties.[7] In particular, several of these alloys have excellent cryogenic mechanical properties as well as interesting electrical and thermal transport properties and increased resistance to radiation damage,[8] providing clear evidence that these unusual properties depend on both the number of elements and specific elemental types presented. In addition, recent experimental results have shown that intrinsic chemical disorder can influence defect dynamics at early stages, and a suppressed radiation damage accumulation is revealed with increasing chemical disorder from pure Ni to binary and even more complex quaternary solid-solution alloys.[8]

The properties of concentrated solid-solutions are closely related to the complexity of the underlying compositionally disordered state. The compositional disorder results in two distinct features that make these alloys fundamentally different from conventional structural alloys. Firstly, the random arrangement of different elements in solid-solutions leads to on-site elemental species (atomic) disorder. Secondly, the random arrangement of surrounding atoms brings about random local distortion of the lattice (displacement fluctuations), in which the atom occupying a particular site is slightly displaced from its ideal lattice position. While atomic and displacement disorders dominate the intrinsic physical properties of these alloys, they also make the energetics of defect formation and migration fundamentally different from conventional pure materials and dilute solid-solutions. For example, the formation and migration energies of vacancies and interstitials take on a distribution of values, rather than having a single value as is generally assumed in dilute alloys. Perhaps more importantly, there is a distribution for migration barriers that has the potential to change the nature of vacancy and interstitial diffusion mechanisms. Furthermore, the possibility of overlap between the distributions of vacancy and interstitial migration barriers determines the degree to which interstitial and vacancy populations can separate. Given that the production and migration of point defects under irradiation directly affects the number, distribution and agglomeration of surviving defects and consequently microstructure and mechanical properties, it is important to develop a quantitative understanding of the properties of point defects in concentrated solid-solutions. Until now, most studies of point defects in alloys have emphasized dilute alloys with low concentrations of solute elements.[9, 10] Comparatively little is known about the defect properties in concentrated solid-solution alloys.

Nowadays, *ab initio* calculations based on density functional theory (DFT) are a standard method for studying the formation and diffusion energy of point defects in metals and dilute alloys.[9-14] However, for concentrated solid-solution alloys, the chemical disorder makes it difficult to calculate the defect energies. Usually, there are two methods that can be used to address this problem. One is the coherent potential approximation (CPA) method that properly takes into account the effects of chemical disorder on the electronic structure of alloys, in particular the configurationally averaged single-site charge density, magnetization density, density of states and total energy.[15] Due to its single-site, or mean-field nature, the CPA does not allow consideration of the effects of local lattice relaxation or displacement fluctuations. Although defects can be modeled as an additional (dilute) species, the distribution of formation energies and migration energies that results from local environmental effects cannot be obtained. The non-local CPA[16] could in principle be developed to include some of these effects. However, to date this has not been done. Additionally, the treatment of migration barriers lies well outside the domain of CPA-based approaches. The more conventional approach is to use supercell models which enables consideration of local lattice distortion, albeit at the expense of replacing the infinite disordered system with a finite supercell that is periodically reproduced and in most instances, completely ignores the effects of configurational averaging. Despite these drawbacks, supercell models do provide detailed information about the defect structure and thus are suitable for describing the distribution of defect energies. Indeed, a recent study of the formation energies of intrinsic point defects in a Fe-10Ni-20Cr model alloy reveals that the variables describing the local environment surrounding a point defect can be used to fit the formation energies of defects.[12] The dominant factor is shown to be the first nearest neighbors around the defect site. However, the supercell size is limited in this method.

Besides the local relaxation, the formation energy of a point defect, such as a vacancy or interstitial, also depends sensitively on the chemical potential of the element being removed or added, which is defined as the Gibbs free energy per atom of the reservoir. In actuality, it is a measure of the free energy variation after the defect is introduced into the alloy lattice. High chemical potential implies that it requires more energy to put the atom back into the alloy and results in higher formation energy of vacancy. In previous studies of defect physics in metals and dilute alloys, the chemical potential of the reservoir is usually taken as the energy per atom of the corresponding elemental solid. However, it is difficult to apply this method to concentrated solid-solution alloys in which elements are at or near an equiatomic ratio. Indeed, the elemental chemical potential in concentrated solid-solution alloys can be much different from that in pure metals. In fact, its determination is a big challenge because of the random arrangement of the atomic species. Therefore, it is desirable to develop an efficient approach to calculate the elemental chemical potential in concentrated solid-solution alloys.

Although DFT calculations are currently the most accurate method to determine the properties of point defects, they can only tackle relatively small systems, typically a few hundred atoms owing to the high computational cost. Thus it is also important to develop classical molecular dynamics (MD) and Monte Carlo (MC) methods based on embedded atom method (EAM) potentials to study the long-term evolution of defects. Unfortunately, their results depend sensitively on the potential used in the calculations. Thus it is critical to validate their accuracy, particularly for the point defect properties they predict. In addition, comparing the results obtained from DFT and EAM method can help to evaluate the performance of EAM potentials and provide an overall picture of defect energetics from different theoretical levels. The comparison is especially important for concentrated solid-solutions since the defect energies exhibit a distribution rather a single value. It is also a necessary step toward multiscale modeling, for which the ultimate goal is to interpret and finally predict the macroscopic behavior of materials.

In this work, the point defect formation and migration energies in $Ni_{0.5}Co_{0.5}$, $Ni_{0.5}Fe_{0.5}$, $Ni_{0.8}Fe_{0.2}$, and $Ni_{0.8}Cr_{0.2}$ concentrated solid-solution alloys have been studied, including vacancies and [100] dumbbell interstitials. It should be noted that these solid-solution alloys remain *fcc* at low temperatures. The high quality of the crystals was confirmed with ion channeling techniques previously.[8] Therefore, the discussion of defect properties in these solid-solution alloys at 0K based on DFT can be used to analyze the effect of extreme disorders. These four alloys are chosen in order to investigate the effect of different compositions and elements on the defect energetics in concentrated solid-solutions. The rest of the paper is organized as follows. Section II describes our computational methods. Section III discusses the results, including the properties of these four alloys in the defect-free state, the calculation of chemical potential used in defect calculations, the calculated formation and migration energies in these four alloys and the comparison between *ab initio* calculations and classical simulations. Finally, Sec. IV gives a summary of the paper.

# 2 Methodology

The alloy structures were modeled utilizing special quasirandom structures (SQS),[17, 18] which were generated by a Monte Carlo algorithm so that the Warren-Cowley short range order parameters[19] in the first four shells approached zero as closely as possible. Here the short range order parameters for shell *i* are defined as

$$\alpha^i = 1 - \frac{1}{2}\frac{N_{AB}^i}{c_i c_j N_{pairs}^i},$$

(1)

where $c_i$ and $c_j$ are the concentration of species *i* and *j*, $N_{AB}^i$ is the number of nearest AB pairs in shell *i* and $N_{pairs}^i$ is the total number of nearest pairs in the same shell. The SQS method is designed to model fully random alloys with relatively small supercells. In this study, 108-atom supercells were generated for each alloy composition in a 3×3×3 *fcc* lattice.

First-principles calculations were performed using the projector augmented wave (PAW) method[20] as implemented in the Vienna *ab initio* simulation package (VASP) code.[21] The Brillouin zone was sampled with a 4×4×4 Γ-centered mesh with an energy cutoff for the plan-wave basis set of 500 eV. First-order Methfessel and Paxton smearing[22] of the Fermi surface was used, with a smearing width of 0.1 eV. The exchange-correlation functional was described by generalized gradient approximations (GGA) in the standard Perdew-Burke-Ernzerhof formalism.[23] Note that GGA is necessary to correctly predict the ground state of Fe to be ferromagnetic (FM) and *bcc*.[24] Standard PAW pseudopotentials supplied with VASP were used, with 16, 14 and 12 valence electrons used for Ni, Fe and Cr, respectively. In this way, the semi-core *p* states were included as valence states in the calculations. Spin-polarized calculations were carried out throughout this work. The magnetic moments of atoms were initialized with the desired magnetic ordering and then allowed to relax during the calculations. The final local magnetic moments were determined by integrating the spin density within spheres centered on the atoms using the default sphere radii. The supercell approach with periodic boundary conditions was used.

For each structure, the internal coordinates of all atoms were first relaxed at a given volume and initial magnetic configuration until the magnitude of the force acting on each atom was smaller than 0.01 eV/Å and the energy difference between two consecutive steps was less than $1\times10^{-6}$ eV. The optimal lattice parameters and bulk moduli were obtained by fitting the calculated energy-volume relation to the Murnaghan equation of state.[25] All of the following defect properties calculations were carried out under this optimal lattice by relaxing only the atomic coordinates with the same convergence criteria. The method of conjugate gradient energy minimization was employed.

The defect formation energy $E_f(\alpha)$ per defect is calculated as[9, 12, 26-28]

$$E_f(\alpha) = E_T(\alpha) - E_0 \pm \mu_\alpha ,$$

(2)

where $E_T(\alpha)$ is the total energy of the supercell with defects $\alpha$, $E_0$ is the total energy of the perfect supercell and $\mu_\alpha$ is the chemical potential of the defect species that added to (-) or removed (+) from the perfect lattice to create the defect. The chemical potential can be varied because of the specific equilibrium growth conditions in the experiment and represents the energy for putting the vacated atom back into the alloy. In this work, these chemical potentials were calculated following Widom-type substitution techniques.[29] Specifically, we randomly substituted the atom species in an alloy with the other species and calculated the energy difference, which is the difference between the chemical potential of these two species:

$$\mu_A - \mu_B = E^{A \to B} - E_0 ,$$

(3)

$$\mu_B - \mu_A = E^{B \to A} - E_0 ,$$

(4)

where $E^{A \to B}$ is the total energy of the alloy with one B atom substituted by A. These two equations are actually the same, and they only provide a criterion for checking our results. Combined with another equation

$$E_0 = N_A \mu_A + N_B \mu_B ,$$

(5)

where $N_A$ and $N_B$ are the numbers of A and B atoms in a perfect lattice, eq. (3) and eq. (4) can be used to solve the chemical potential for these atom species. In practical calculations, we merely performed several sets of substitutions and averaged their energies to feed into the eq. (3), eq. (4) and eq. (5). More details about the determination of chemical potential are provided in the next section.

The mixing enthalpy of the alloy was defined as $H_{mix} = H_{A_x B_{1-x}} - xH_A - (1-x)H_B$, where $H_{A_x B_{1-x}}$, $H_A$, and $H_B$ are the total enthalpy of the $A_x B_{1-x}$ alloy and the corresponding pure metals A and B, respectively.

The diffusion barriers and paths were investigated using the climbing-image nudged elastic band (CI-NEB) method.[30] In this method, a number of intermediate images are optimized along the reaction path. Three to five intermediate images were used, and the energy barriers were determined until the forces on each image were converged to 0.03 eV/Å. Because of the computational cost of NEB calculations, a 2×2×2 K-point sampling was used in this case.

To evaluate the performance of the available EAM potentials in predicting the defect properties of Ni-Fe alloys, we adopted the Large-scale Atomic/Molecular Massively Parallel Simulator (LAMMPS) code.[31] The formation energies of defects were calculated by directly minimizing the energy and force of each configuration using a conjugate gradient method. The convergence criterion was set so that the change in energy between minimization iterations was $1\times10^{-6}$ and the force was below $1\times10^{-10}$ eV/Å. Three Ni-Fe potentials based on the embedded atom method (EAM) developed by Bonny et al. were considered.[32-34] The diffusion barrier of the defects was determined by the CI-NEB method as implemented in LAMMPS, using the "quick-min" damped minimization algorithm with a time step of 0.01 fs. The NEB calculations were stopped when the force on each atom was less than $1\times10^{-6}$ eV/Å. The following CI-NEB calculations proceeded with the same stopping criterion, and 16 intermediate images were used.

# 3 Results

## 3.1 Bulk materials

The structural properties of bulk *fcc* Ni, *bcc* Fe, and *bcc* Cr are first investigated. For Co, we choose the hexagonal close packed (*hcp*) structure since it is the equilibrium crystalline structure at room temperature.[35] The results are summarized in Table. I. The calculated lattice parameters, the bulk moduli, and the magnetic ground states for these four metals are in good agreement with previous results. Note that the ground states of Ni, Co and Fe are FM whereas Cr is antiferromagnetic (AFM).

TABLE I. Calculated properties of bulk Ni, Co, Fe and Cr metals: lattice parameter $a_0$ (Å), bulk modulus $B_0$ (GPa), magnetic moment μ (μ$_B$/atom), formation energy $E_f(V)$ (eV) and migration energy $E_m(V)$ (eV) of vacancies as well as the formation and migration energy of interstitials $E_f(I)$ and $E_m(I)$ (eV).

|    |         | $a_0$ | $B_0$ | μ | $E_f(V)$ | $E_m(V)$ | $E_f(I)$ | $E_m(I)$ |
|----|---------|-------|-------|---|----------|----------|----------|----------|
| Ni | Present | 3.527 | 178.7 | 0.644 | 1.47 | 1.01 | 4.27$^a$ | 0.11 |
|    | Theo.   | 3.518[36] | 198[37] | 0.61[38] | 1.43[39],1.46[37] | 1.08[39] | 4.07[39] | 0.14[39] |
|  | Expt. | 3.5240[40] | 181[41] | 0.61[42] | 1.58-1.63[43],1.8[44] | 1.27[43],1.1[44] |  |  |
| Co | Present | 2.486 | 212 | 1.60 | 1.90 |  |  |  |
|  | Theo. | 2.476[45],2.5007[35] | 221[46] | 1.58[35],1.60[46] |  |  |  |  |
|  | Expt. | 2.507[40] | 191.4[42] |  | 1.34[47] |  |  |  |
| Fe | Present | 2.840 | 164.9 | 2.249 | 2.23 |  |  |  |
|  | Theo. | 2.833[36],2.85[48] | 160[13],195[14] | 2.32[13],2.24[48] | 2.10[48],2.15[49] | 0.65[13] | 3.64[50] | 0.34[51] |
|  | Expt. | 2.8665[40] | 166.2[52] | 2.22[42] | 2.0[53] | 0.55[54] |  |  |
| Cr | Present | 2.870 | 177.6 | 0.976 | 2.76$^b$,2.88$^c$ |  |  |  |
|  | Theo. | 2.849[55] | 189[55] | 0.92[55] | 2.85[14],2.71[49] |  |  |  |
|  | Expt. | 2.8848[40] | 191[56] | 0.60[42] | 2.0±0.2[57] |  |  |  |

$^a$ [100] dumbbell.
$^b$ FM
$^c$ AFM

Some intrinsic point defects in these four metals are also studied. While vacancy defects are relatively simple, there exist several interstitial sites in these metals. Since we are interested in Ni-based *fcc* solid-solutions, we focus on the *fcc* structure of Ni. In pure Ni, there are six possible interstitial configurations: octahedral, tetrahedral, and crowdion, as well as [100], [110] and [111] dumbbells. Among them, our calculations show that the [100] dumbbell has the lowest formation energy and is the most stable interstitial configuration, in line with previous reports.[39] The formation energy is calculated to be 4.27 eV, comparable with the previous data.[39]

The diffusion of vacancies in *fcc* Ni is achieved by exchanging an atom with a nearest neighbor. The calculated barrier for this process is 1.01 eV, in good agreement with the previous result.[39] The most preferable diffusion pathway for a [100] dumbbell interstitial in *fcc* Ni is to convert to a [010] dumbbell by a shifting and rotation mechanism.[39] The energy barrier for this path is determined to be 0.11 eV in this work by the energy difference between the highest energy and the lowest energy along the reaction coordinates. In this process, the energy first goes down by −0.10 eV and then up to 0.01 eV at the middle of the path. The diffusion energy profile has two symmetric low energy valleys.

The structural properties of the four alloys considered in this work are presented in Table II. It is notable that the bulk moduli of the binary alloys are larger than those of the constituent pure metals. For example, the bulk modulus of $Ni_{0.5}Fe_{0.5}$ is calculated as 186.9 GPa, larger than the 178.7 GPa of pure Ni and 164.9 GPa of pure Fe. This observation suggests that concentrated solid-solution alloys can be stronger than pure metals. The mixing enthalpy of $Ni_{0.5}Co_{0.5}$ is positive as the *hcp* Co is used as the reference state. The mixing enthalpy of $Ni_{0.8}Fe_{0.2}$ is lower than that of $Ni_{0.5}Fe_{0.5}$, suggesting that $Ni_{0.8}Fe_{0.2}$ is the preferable phase for a Ni-Fe alloy compared with $Ni_{0.5}Fe_{0.5}$. This is because $Ni_{0.8}Fe_{0.2}$ is closer to $L1_2$-$Ni_3$Fe phase, which is the most stable phase for a Ni-Fe alloy.[58, 59] $Ni_{0.8}Cr_{0.2}$ has a positive mixing energy, which is an indication of a possible phase separation for Ni and Cr. In fact, a solid-solution of Cr with Ni is not stable in the *fcc* phase when the concentration of Cr increases.

TABLE II. Calculated properties of the AB alloys considered in this work: lattice parameter $a_0$ (Å), bulk modulus $B_0$ (GPa), mixing enthalpy $H_{mix}$ per formula unit (f.u.) (meV/f.u.) and magnetic moment $\mu_A$ and $\mu_B$ ($\mu_B$/atom), respectively.

| Alloy | $a_0$ | $B_0$ | $H_{mix}$ | $\mu_A$ | $\mu_B$ |
|---|---|---|---|---|---|
| $Ni_{0.5}Co_{0.5}$ | 3.525 | 204.7 | 21.5 | 0.650 | 1.710 |
| $Ni_{0.5}Fe_{0.5}$ | 3.579 | 186.9 | −34.1 | 0.708 | 2.667 |
| $Ni_{0.8}Fe_{0.2}$ | 3.546 | 191.8 | −80.7 | 0.680 | 2.772 |
| $Ni_{0.8}Cr_{0.2}$ | 3.531 | 191.7 | 43.9 | −0.179 | 0.021 |

The calculated ground states of $Ni_{0.5}Co_{0.5}$, $Ni_{0.5}Fe_{0.5}$ and $Ni_{0.8}Fe_{0.2}$ are ferromagnetic with Ni, Co and Fe local moments aligned in the same direction. For $Ni_{0.8}Cr_{0.2}$, the case is more complicated owing to the fact that the Cr ground state is itself AFM and it couples AFM (anti-aligned) with respect to Ni. We compare the energies of optimized structures of $Ni_{0.8}Cr_{0.2}$ initiated by different spin. In the first structure, all the atomic moments are in parallel directions; in the second, all the magnetic moments are parallel for Ni atoms, whereas the moments of Cr atoms are in alternately parallel and antiparallel directions; and in the last, all the magnetic moments are in alternately parallel and antiparallel directions. The total energies obtained from these three configurations are different, with the third one having the lowest energy (about 0.27 eV lower than the first one). The energies of the second and third structure are very similar, with an energy difference of 0.04 eV. Therefore, we use the third structure as the ground state of the $Ni_{0.8}Cr_{0.2}$ structure. In this case, most magnetic moments of the Ni atoms are parallel to one another, whereas the moments of Cr atoms are alternately parallel and antiparallel, suggesting that Cr atoms tend to be AFM. The local moments carried by each atom species are summarized in Table II. Note that because of the AFM state of Cr, the averaged value is small (0.021 $\mu_B$/atom). Actually, the averaged absolute moment of Cr is 1.34 $\mu_B$/atom, which is much larger than 0.179 $\mu_B$/atom of Ni.

The atomic displacement of these four alloys is obtained after the atomic coordinates are fully relaxed. The averaged displacement in $Ni_{0.5}Co_{0.5}$ is -0.0045Å for the Ni-Ni first neighbor pairs, −0.0014 Å for the Ni-Co first neighbor pairs and 0.0071 Å for the Co-Co first neighbor pairs. The magnitude of these displacements is small, indicating less distortion in $Ni_{0.5}Co_{0.5}$ solid solution alloys. This is reasonable as Co can form *fcc* structure with the lattice constant similar to that of *fcc* Ni. For $Ni_{0.5}Fe_{0.5}$, it is found that the averaged displacement is -0.0022 Å for the Ni-Ni first neighbor pairs, −0.0095 Å for the Ni-Fe first neighbor pairs and 0.0226 Å for the Fe-Fe first neighbor pairs. The large displacement of Fe-Fe and the contraction of the Ni-Fe distance are in accordance with previous experimental data,[60] although the atomic composition is not exactly the same. The displacement of Fe-Fe is the largest, even for the second and the third neighbors. This is also the case for the $Ni_{0.8}Fe_{0.2}$ alloy. These results indicate that FM Fe has a large volume in Ni, which will affect the defect properties in these alloys. For $Ni_{0.8}Cr_{0.2}$, the largest displacement is observed in Cr-Cr for all the neighbor shells. The first neighbor of Cr-Cr has an expansion of 0.0089 Å, while the second neighbor shells contracts by −0.0022 Å. This observation is in accordance with previous conclusions that Cr is oversized in the Ni matrix.[61] Note that the displacement of atoms in solid-solutions plays an important role in the defect energetics, since it will change the local environment around the defect site and lead to different lattice relaxations.

## 3.2 Defect properties

Owing to the chemical disorder, the defect configurations are difficult to generalize in a totally random alloy. Therefore, only the vacancies and [100] dumbbells are considered.

### *3.2.1 Calculation of chemical potentials*

The chemical potentials of the elements in each alloy are calculated by substituting the atoms with the other species in the corresponding alloy structure. For each element we carry out a series of substitution energy calculations according to different first nearest neighbor shells. In a perfect *fcc* lattice, there are 12 atoms in the first neighbor shell. Consequently, we denote the first neighbor pair as (*m,n*), where *m* is the number of Ni atoms and *n* is the number of Fe or Cr atoms in the first nearest neighbor. For those atoms with the same (*m,n*), we randomly choose one to perform the substitution. Because of the finite supercell size, not all combinations of (*m,n*) can be found. Therefore, the (*m,n*) considered should conform to the concentration ratio after averaging. For example, the (*m,n*) used in $Ni_{0.8}Cr_{0.2}$ have to be pairs such as (9,3) and (10,2), which gives an average ratio of 19:5 that is closest to the concentration ratio 4:1. Then Ni or Cr atoms with different nearest neighbor pairs are chosen to perform the substitution calculation. The substitution energy is averaged from these calculations and validated by a cross-check with eq. (3) and (4). Using this method, our calculations show good agreement with the two equations, and the absolute energy difference between $\mu_A - \mu_B$ and $\mu_B - \mu_A$ is rather small. The difference is only 0.01 eV for $Ni_{0.5}Co_{0.5}$, 0.02 eV for $Ni_{0.5}Fe_{0.5}$, 0.00 eV for $Ni_{0.8}Fe_{0.2}$ and 0.01 eV for $Ni_{0.8}Cr_{0.2}$.

Usually, calculating the chemical potential in a random alloy is a big challenge, since sufficient substitutions should be carried out to obtain an accurate average value. By classifying different atoms in the alloy according to their nearest neighbors, we show that the averaging can be performed efficiently only within those atoms surrounded by different nearest neighbor pairs. In actuality, this method preserves the concentration of the alloy and thus can obtain a good sampling of the substitution energies. The method can also be applied to triple or multi-component solid-solution alloys to obtain accurate chemical potentials. The small difference between $\mu_A - \mu_B$ and $\mu_B - \mu_A$ shown above in $Ni_{0.5}Co_{0.5}$, $Ni_{0.5}Fe_{0.5}$, $Ni_{0.8}Fe_{0.2}$, and $Ni_{0.8}Cr_{0.2}$ indicates that our substitution method provides an approach of effective sampling, and greatly facilitates the calculation of elemental chemical potentials in random alloys.

Here it is helpful to compare the calculated chemical potentials in these alloys with those obtained directly from their pure metal reference states. For $Ni_{0.5}Co_{0.5}$, the difference is 0.01 eV for both Ni and Co. The difference between Ni and Fe is −0.08 and 0.05 eV in $Ni_{0.5}Fe_{0.5}$, and −0.03 and −0.07 eV in $Ni_{0.8}Fe_{0.2}$. The lower elemental chemical potential in the alloys suggests that the element is more stable in the corresponding alloys than in the pure metal phase. Thus, Ni is energy-preferable for forming an *fcc* solid-solution with Fe. The chemical potential difference of Ni is close to zero going from $Ni_{0.5}Fe_{0.5}$ to $Ni_{0.8}Fe_{0.2}$, indicating it is less costly to put Ni back into the Ni-rich $Ni_{0.8}Fe_{0.2}$ alloy. The lower chemical potential of Fe in $Ni_{0.8}Fe_{0.2}$ than in $Ni_{0.5}Fe_{0.5}$ is in agreement with the much lower mixing energy of $Ni_{0.8}Fe_{0.2}$. For $Ni_{0.8}Cr_{0.2}$, it is found that the chemical potential of Ni is −0.01 eV smaller in $Ni_{0.8}Cr_{0.2}$ than in bulk Ni metal, whereas the chemical potential of Cr is 0.14 eV larger in $Ni_{0.8}Cr_{0.2}$ than in bulk Cr metal. This result provides further evidence of the low affinity of Cr to Ni.

### *3.2.2 Vacancy*

Vacancies are common defects in bulk materials. The monovacancy formation energies in the four alloys considered in this work are presented in Fig. 1. In a random alloy, the formation energy of a vacancy is dependent on its environment. Therefore, we classify these formation energies according to the first nearest neighbor atoms around the vacancy.

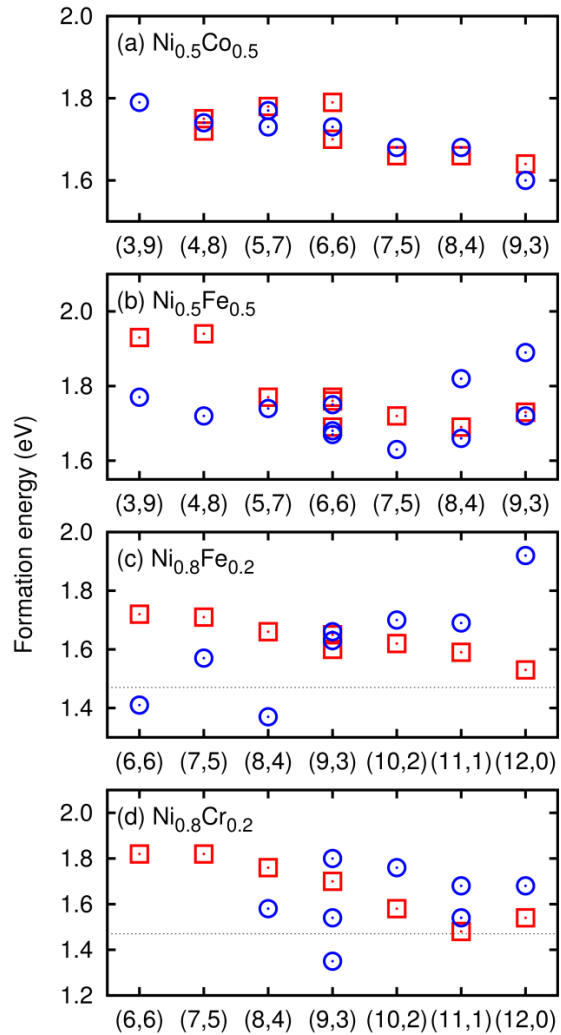

Fig. 1 (Color online) Dependence of vacancy formation energies (in eV) in four Ni-based alloys on the number of its nearest neighbor pair (*m*, *n*), where *m* is the number of Ni and *n* is the number of Fe or Cr: (a) $Ni_{0.5}Co_{0.5}$, (b) $Ni_{0.5}Fe_{0.5}$, (c) $Ni_{0.8}Fe_{0.2}$ and (d) $Ni_{0.8}Cr_{0.2}$. A Ni vacancy is denoted by an empty square and a Co, Fe or Cr vacancy is an empty circle. The dotted line in (c) and (d) represents the vacancy formation energy in pure Ni (1.47 eV).

It can be seen from Fig. 1(a) that the formation energy of Ni and Co vacancies in $Ni_{0.5}Co_{0.5}$ are comparable, exhibiting the same trend. This is also a result from the less distortion observed in this alloy. The formation energy of Ni vacancy in $Ni_{0.5}Fe_{0.5}$ tends to decrease as nearest Ni neighbors increase (Fig.1(b)). This trend is more prominent for the $Ni_{0.8}Fe_{0.2}$ shown in Fig. 1(c). The formation energies of Fe vacancies seem to rely greatly on the local environment, but there is an increasing tendency with the increase of Ni neighbors. These results indicate that Ni vacancies prefer a Ni-rich local environment and Fe vacancies prefer a Fe-rich environment. High Ni-coordinated Ni sites are much more likely to be vacated in a Ni-rich $Ni_{0.8}Fe_{0.2}$ alloy. The concentration of vacancies in $Ni_{0.8}Fe_{0.2}$ will be much larger than in $Ni_{0.5}Fe_{0.5}$ because of their lower formation energies. Compared with vacancy formation energies in pure Ni and Fe metals (1.47 and 2.23 eV), we find that all the calculated vacancy formation energies in $Ni_{0.5}Fe_{0.5}$ and $Ni_{0.8}Fe_{0.2}$ fall between the corresponding vacancy formation energies of pure Ni and pure Fe.

The local magnetic moments of alloys are affected by introducing vacancies. For $Ni_{0.5}Co_{0.5}$ and $Ni_{0.5}Fe_{0.5}$, our calculations show that the magnetic moments of both Ni and Fe atoms in the first nearest neighbor position of the vacancy site increase. In $Ni_{0.5}Co_{0.5}$, the increase is 0.016 $\mu_B$/atom and 0.038 $\mu_B$/atom for nearest Ni and Co atoms around Ni vacancies while it is 0.012 $\mu_B$/atom and 0.050 $\mu_B$/atom around Co vacancies. In $Ni_{0.5}Fe_{0.5}$, for Ni vacancies, the averaged magnitude of increase is 0.011 $\mu_B$/atom for the nearest Ni and 0.042 $\mu_B$/atom for the nearest Fe atoms. The increase is 0.005 $\mu_B$/atom for the nearest Ni and 0.074 $\mu_B$/atom for the nearest Fe atoms around Fe vacancies. The results reflect the fact that the vacancy makes more space for the relaxation of the magnetic moments. The averaged variation of magnetic moments of the nearest Ni and Fe atoms is 0.011 and 0.043 $\mu_B$/atom for Ni vacancies and 0.007 and 0.072 $\mu_B$/atom for Fe vacancies in $Ni_{0.8}Fe_{0.2}$. The variation of moments $Ni_{0.5}Fe_{0.5}$ and $Ni_{0.8}Fe_{0.2}$ is similar.

The atomic structural relaxation around a vacancy site is calculated by comparing the atomic coordinates in the optimized perfect lattice and the vacant one. For $Ni_{0.5}Co_{0.5}$, the relaxation of Ni and Co first-nearest neighbor shell around vacancies is very similar. Actually, their averaged relaxations are all about −0.010 Å around Ni and Co vacancies. This is also related to the very small distorted *fcc* lattice in $Ni_{0.5}Co_{0.5}$. In $Ni_{0.5}Fe_{0.5}$, it is found that the Fe first-nearest neighbor shell undergoes the largest inward relaxation for both Ni and Fe vacancies. The result is −0.020 Å for the Fe first nearest-neighbor compared with −0.002 Å for the Ni first-nearest neighbor around Ni vacancies, and −0.029 Å compared with −0.004 Å around Fe vacancies, respectively. The relaxation becomes very small for second nearest neighbors. That is also the case for $Ni_{0.8}Fe_{0.2}$, in which the relaxation of the Fe first-nearest neighbor is −0.018 Å, compared with −0.009 Å for Ni vacancies, and −0.026 Å compared with −0.008 Å for Fe vacancies. The larger relaxation of Fe neighbors contributes to the larger variation of their magnetic moments.

For $Ni_{0.8}Cr_{0.2}$, the formation energy of Ni vacancies decreases as Ni in the nearest neighbor shell increases. The relaxation of the nearest Cr neighbors around vacancies is larger than the Ni neighbors. For a Ni vacancy, the averaged relaxation of Cr neighbors is −0.018 Å whereas it is

–0.013 Å for the Ni neighbors. For Cr vacancies, the local relaxation averages –0.021 Å for Cr neighbors and –0.011 Å for Ni neighbors. The antiparallel direction of magnetic moment for Cr atoms is preserved after a vacancy is introduced. However, the change in magnetic moments for the nearest neighbors around vacancies is somewhat random. When all the vacancies investigated are averaged, the respective variation in magnetic moments for the nearest Ni and Cr atoms is 0.00 and –0.02 $\mu_B$/atom, respectively, around Ni vacancies, –0.05 and –0.18 $\mu_B$/atom, respectively, around Cr vacancies. Again, the larger relaxation is in accordance with the larger variation in magnetic moments of Cr atoms around Cr vacancies.

By comparing these four alloys, we see that the formation energy of vacancies depends on both the composition and the local environment. This is reflected in the formula of eq. (2), which shows that formation energy consists of two parts: one is the energy to remove the atom and the other is the chemical potential of the atom. While the former is determined by the local environment, the latter is composition dependent. Most of the formation energies of vacancies in these solid-solutions are larger than those in pure Ni (1.47 eV), which suggests that solid-solutions help to decrease the vacancy concentrations compared with pure Ni and thus make these alloys more resistant to vacancy formation. In this sense, adding Fe into the Ni matrix is an efficient way to suppress vacancy formation.

The diffusion barrier is an important parameter in transport properties. The diffusion barriers of vacancies are determined by NEB calculations. We mainly consider the diffusion path by directly exchanging atoms between first nearest neighbors. The results are presented in Table III. The initial and final states are denoted by the number of atoms in the first nearest neighbor (*m*,*n*) around vacancies. The forward ($E_m^f$) and reverse ($E_m^r$) energy barriers are given, which are defined as the energy difference between the highest replica and the first or last replica, since the formation energy of vacancies is dependent on their environment.

TABLE III. The forward ($E_m^f$) and reverse ($E_m^r$) migration energy barriers (in eV) of vacancies in four alloys

| Alloy | Vacancy | Path | $E_m^f$ | $E_m^r$ |
|---|---|---|---|---|
| Ni$_{0.5}$Co$_{0.5}$ | Ni | (6,6)–(4,8) | 1.21 | 1.25 |
| | | (6,6)–(5,7) | 1.16 | 1.17 |
| | | (6,6)–(6,6) | 1.24 | 1.15 |
| | | (8,4)–(9,3) | 1.19 | 1.17 |
| | Co | (6,6)–(3,9) | 1.16 | 1.10 |
| | | (6,6)–(7,5) | 1.09 | 1.14 |
| | | (6,6)–(5,7) | 1.15 | 1.11 |
| | | (3,9)–(7,5) | 1.04 | 1.15 |
| Ni$_{0.5}$Fe$_{0.5}$ | Ni | (6,6)–(4,8) | 1.48 | 1.31 |
| | | (6,6)–(6,6) | 1.2 | 1.19 |
| | | (6,6)–(9,3) | 1.33 | 1.37 |
| | | (9,3)–(5,7) | 1.22 | 0.98 |
| | Fe | (6,6)–(5,7) | 0.97 | 0.93 |
| | | (6,6)–(6,6) | 0.94 | 1.01 |
| | | (6,6)–(7,5) | 0.97 | 1.17 |
| | | (8,4)–(5,7) | 1.07 | 1.15 |
| $Ni_{0.8}Fe_{0.2}$ | Ni | (9,3)–(8,4) | 1.18 | 1.17 |
| | | (9,3)–(9,3) | 1.15 | 1.19 |
| | | (9,3)–(10,2) | 1.31 | 1.34 |
| | | (11,1)–(10,2) | 1.35 | 1.32 |
| | Fe | (9,3)–(6,6) | 0.82 | 1.04 |
| | | (9,3)–(7,5) | 1.04 | 0.95 |
| | | (9,3)–(9,3) | 1.04 | 1.01 |
| | | (10,2)–(8,4) | 0.95 | 1.28 |
| $Ni_{0.8}Cr_{0.2}$ | Ni | (9,3)–(7,5) | 1.05 | 0.98 |
| | | (9,3)–(8,4) | 1.13 | 1.24 |
| | | (9,3)–(10,2) | 0.96 | 1.01 |
| | | (6,6)–(8,4) | 1.01 | 1.14 |
| | Cr | (9,3)–(8,4) | 0.68 | 0.9 |
| | | (9,3)–(9,3) | 0.9 | 0.65 |
| | | (11,1)–(8,4) | 1.16 | 1.19 |
| | | (11,1)–(9,3) | 0.86 | 0.72 |

It can be seen that in $Ni_{0.5}Co_{0.5}$, all the calculated migration barriers are higher than that in pure Ni (1.01 eV), indicating slower diffusion for vacancies. For both $Ni_{0.5}Fe_{0.5}$ and $Ni_{0.8}Fe_{0.2}$, the diffusion barrier for Ni vacancies is larger than in pure Ni, while the barrier for Fe vacancies is smaller. The low diffusion barrier of Fe vacancies and high diffusion barrier of Ni vacancies suggests that Fe vacancies are more mobile than Ni vacancies in Ni-Fe solid-solutions. Nevertheless, the diffusion barrier is dependent on the local environment of the vacancy.

In $Ni_{0.8}Cr_{0.2}$, the vacancy diffusion relies strongly on the location of the vacancy. The migration energy is around 0.7–1.2 eV for the specific diffusion path we have investigated. Compared with pure Ni, we can see that the incorporation of Cr does not induce a significant difference in the diffusion of Ni vacancies in $Ni_{0.8}Cr_{0.2}$.

The diffusion of vacancies is equivalent to the migration of a lattice atom between two nearest neighbor sites. In a pure metal, the total energy of the system increases smoothly during migration from the starting point to the middle point and then decreases to the end point. The highest energy barrier is observed when the atom passes through the middle point between the two nearest neighbors. However, the atomic displacement presented in solid-solutions greatly changes the diffusion kinetics owing to different displacements at different atomic species sites. Thus the relaxation induced by different vacancies is significantly different. As a result, the reaction coordinates for vacancy diffusion depend on the local environment around the vacancy site, and the migration barrier is not always located at the middle point. This effect is less pronounced for vacancy diffusion than for interstitial diffusion, as the introduction of interstitials will result in a large relaxation of the lattice.

### 3.2.3 [100] dumbbell

The most stable interstitial defects in metals such as Ni are dumbbells with two atoms sharing a lattice site. We investigated the formation and migration energies of [100] dumbbells in the four alloys, as this dumbbell is the most stable interstitial defect in an *fcc* Ni structure. In general, the formation of interstitials leads to large relaxations of the lattice. Thus, we have analyzed the effect of a finite supercell size on the formation energies of interstitials. If the supercell is too small, it will induce additional interactions between an interstitial atom and its images due to periodic boundary conditions. Consequently, the calculated formation energies from small supercells will be overestimated. As the formation energies in these concentrated alloys with randomly arranged elements are distribution rather than a single value, it is not possible to make a direct comparison unless the whole distribution is obtained. Therefore, we chose the case of pure Ni as an example to illustrate the error induced by the finite supercell. The formation energy of a [100] dumbbell in a 108-atom supercell is 4.27 eV, while a formation energy of 4.21 eV is obtained for a 256-atom supercell. The difference is 0.06 eV suggesting the error of calculated formation energies for interstitials is only 1%, which is small. Thus our results obtained from a finite supercell size will be slightly overestimated, but the relative trend of the formation energy among these alloys should nevertheless be sensible.

The calculated formation energies in the four alloys considered are summarized in Table. IV. The dumbbells are introduced at different lattice sites. We use the number of nearest neighbors around the defect site to represent the local environment.

TABLE IV. Typical formation energies (in eV) of [100] dumbbells at different atomic sites in four alloys

| Alloy | Site | Dumbbell | $E_f(I)$ |
|---|---|---|---|
| $Ni_{0.5}Co_{0.5}$ | (4,8)Ni | [100]Ni-Ni | 4.02 |
| | | [100]Ni-Co | 3.87 |
| | (4,8)Co | [100]Co-Ni | 3.96 |
| | | [100]Co-Co | 3.78 |
| | (6,6)Ni | [100]Ni-Ni | 3.95 |
| | | [100]Ni-Co | 3.81 |
| | (6,6)Co | [100]Co-Co | 3.85 |
| | | [100]Co-Ni | 4.04 |
| | (8,4)Ni | [100]Ni-Ni | 4.01 |
| | | [100]Ni-Co | 3.92 |
| | (8,4)Co | [100]Co-Ni | 3.96 |
| | | [100]Co-Co | 3.79 |
| $Ni_{0.5}Fe_{0.5}$ | (3,9)Ni | [100]Ni-Ni | 3.18 |
| | | [100]Ni-Fe | 3.43 |
| | (3,9)Fe | [100]Fe-Ni | 3.38 |
| | | [100]Fe-Fe | 3.25 |
| | (6,6)Ni | [100]Ni-Ni | 3.55, 3.51 |
| | | [100]Ni-Fe | 3.63, 3.66 |
| | (6,6)Fe | [100]Fe-Fe | 3.61, 3.69 |
| | | [100]Fe-Ni | 3.47, 3.56 |
| | (9,3)Ni | [100]Ni-Ni | 3.55 |
| | | [100]Ni-Fe | 3.09 |
| | (9,3)Fe | [100]Fe-Ni | 3.66 |
| | | [100]Fe-Fe | 3.69 |
| $Ni_{0.8}Fe_{0.2}$ | (8,4)Ni | [100]Ni-Ni | 3.81 |
| | | [100]Ni-Fe | 4.24, 3.85 |
| | (8,4)Fe | [100]Fe-Ni | 3.75 |
| | | [100]Fe-Fe | 4.16 |

|  | (9,3)Ni | [100]Ni-Ni | 4 |
|  |  | [100]Ni-Fe | 4.3 |
|  | (9,3)Fe | [100]Fe-Ni | 4.1 |
|  |  | [100]Fe-Fe | 4.37 |
|  | (10,2)Ni | [100]Ni-Ni | 4.10, 4.01 |
|  |  | [100]Ni-Fe | 4.3 |
|  | (10,2)Fe | [100]Fe-Ni | 4.1 |
|  |  | [100]Fe-Fe | 4.03 |
|  | (11,1)Fe | [100]Fe-Fe | 4.17 |
| $Ni_{0.8}Cr_{0.2}$ | (8,4)Ni | [100]Ni-Ni | 4.24 |
|  |  | [100]Ni-Cr | 3.89 |
|  | (8,4)Cr | [100]Cr-Ni | 3.64 |
|  |  | [100]Cr-Cr | 3.93 |
|  | (9,3)Ni | [100]Ni-Ni | 4.1 |
|  |  | [100]Ni-Cr | 4.06 |
|  | (9,3)Cr | [100]Cr-Ni | 4.03 |
|  |  | [100]Cr-Cr | 4.09 |
|  | (10,2)Ni | [100]Ni-Ni | 4.03 |
|  |  | [100]Ni-Cr | 3.72 |
|  | (10,2)Cr | [100]Cr-Ni | 3.84 |
|  |  | [100]Cr-Cr | 4.04 |

The formation energy of [100] dumbbell defect is the lowest for Co-Co and the highest for Ni-Ni in $Ni_{0.5}Co_{0.5}$, indicating that Co-Co dumbbell is more stable than Ni-Ni. Besides, these formation energies are all lower than that in pure Ni (4.27 eV). In $Ni_{0.5}Fe_{0.5}$, the formation energy of a [100] Fe-Fe dumbbell is larger than that of a Ni-Ni dumbbell at all the considered atomic sites. Thus Fe-Fe dumbbells are unlikely to be formed in the alloy, while a Ni-Ni dumbbell is the most stable. These formation energies are much smaller than the dumbbell formation energy in pure Ni, but they are still high enough that these defects cannot be formed by thermal perturbation, but only by ion irradiation. The lower formation energies compared with that in pure Ni suggest that $Ni_{0.5}Co_{0.5}$ and $Ni_{0.5}Fe_{0.5}$ is more susceptible to irradiation-induced interstitials. Likewise, the formation energies in $Ni_{0.8}Fe_{0.2}$ are largest for Fe-Fe dumbbells. However, those energies are comparable to that in pure Ni. Therefore, the alloying of Co and Fe to Ni in equiatomic ratio tends to decrease the formation energy of dumbbell defects. The lower formation energy of Ni interstitials suggests that interstitial defects will diffuse preferentially in Ni and will not diffuse into the Fe phase.

The formation of dumbbells leads to perturbations in the magnetic properties. In $Ni_{0.5}Co_{0.5}$, the magnetic moments of Ni atoms in dumbbell defects increase, while the moments of Co atoms decrease. For an example, the moment for a (6,6) Ni atom increase from 0.646 to 0.680 $\mu_B$ after the introduction of dumbbell defect. In $Ni_{0.5}Fe_{0.5}$, the local magnetic moments of Ni atoms in dumbbell defect increase, while the moments of Fe decrease and even flip. For an example, the moment of the original Ni atom in a Ni-Ni dumbbell introduced at a (6, 6) Ni atom site in $Ni_{0.5}Fe_{0.5}$ increases from 0.718 to 0.774 $\mu_B$ while the moment of the other Ni atom is 0.762 $\mu_B$. Note that the moments of remaining Ni atoms average 0.704 $\mu_B$. For a Fe-Fe dumbbell at a (6, 6) Fe atom site, the moment of the original Fe decreases from 2.676 $\mu_B$ to 0.210 $\mu_B$, and the moment of the other Fe is 0.107 $\mu_B$; the moments of remaining Fe atoms average 2.623 $\mu_B$. For a mixed dumbbell, the same trend is seen: the moments of Ni atoms increase and those of Fe atoms decrease. The small moment for Fe-Fe dumbbells is also found in the other alloys.[9] In $Ni_{0.8}Fe_{0.2}$, the same tendency in the variation of magnetic moments is observed, although the magnitude of the variation is smaller than in $Ni_{0.5}Fe_{0.5}$. For example, a Fe-Fe dumbbell created at an (8, 4) Fe site leads to a decrease in moments from 2.720 to 1.884 $\mu_B$.

The change in moments is related to the relaxation due to the presence of the dumbbell. Unlike in pure metals, where the atomic displacement is not important, the relaxation of atoms in a solid-solution affects the lattice distortion induced by interstitials. In general, for dumbbell defects, the nearest neighbor atoms residing in the plane perpendicular to the dumbbell axis undergo tensile forces, whereas others undergo compressive forces. However, the forces are not symmetric with respect to the dumbbell axis in solid-solutions because of the inherent atomic displacement (as discussed in the previous section). As a result, the lattice distortion is strongly dependent on the interstitial site. The distinction between tensile and compressive forces does not hold. Therefore, the influence of dumbbell defects on the local magnetic moments is not symmetric as in pure metal[9] and exhibits somewhat random characteristics.

For $Ni_{0.8}Cr_{0.2}$, it is demonstrated in Table IV that a mixed Ni-Cr dumbbell has the smallest formation energies, compared with Ni-Ni and Cr-Cr. Moreover, these energies are smaller than those in pure Ni, indicating that they can easily be created. Taking into account the positive mixing energy for $Ni_{0.8}Cr_{0.2}$ presented in Table II, we see that additional Cr atoms prefer to bind with Ni rather than Cr. Therefore, ion irradiation is unlikely to induce phase separation in $Ni_{0.8}Cr_{0.2}$.

The magnetic coupling is preserved after the introduction of dumbbell defects in $Ni_{0.8}Cr_{0.2}$. The moments of Ni atoms tend to be parallel and those of Cr atoms antiparallel, including the interstitial atoms. For both Ni and Cr interstitials, the variation in the magnetic moments carried by the nearest Cr atoms is larger than that in the nearest Ni atoms. For example, the introduction of a Ni interstitial at a (9, 3) Ni atom site leads to an average variation of 0.018 $\mu_B$/atom and −0.781 $\mu_B$/atom for the moments of the nearest nine Ni and three Cr atoms. The Cr interstitial at the same site results in a 0.070 $\mu_B$/atom and −0.862 $\mu_B$/atom variation for the nearest atoms. The large variation of the nearest Cr atoms is related to the AFM trend of Cr, which induces electron redistribution around the defect site.

Owing to their relatively large formation energies, these interstitials can be produced only by energetic ion bombardment. Under irradiation, the migration of point defects is critical for the occurrence of materials degradation. In this work, we investigated the diffusion mechanism of [100] dumbbells in these solid-solutions. We mainly consider the migration of an interstitial in a [100] dumbbell by making the jump to its nearest-neighbor sites creating a [010] dumbbell, since it is the most preferable pathway in Ni. As a result of the chemical disorder of the alloys, there are three possible migration paths for each atomic species. For example, a Ni interstitial can migrate from a [100] NiNi dumbbell to a [010] NiNi dumbbell, or from a [100] NiNi dumbbell to a [010] NiCr dumbbell, or from a [100] NiCr dumbbell to a [010] NiCr dumbbell. For each diffusion path, we have chosen a specific pathway to investigate the diffusion barriers. The calculated migration energies are summarized in Table V, along with the diffusion paths.

TABLE V. The forward ($E_m^f$) and reverse ($E_m^r$) migration energy barriers (in eV) of interstitials in four alloys

| Alloy | Int. | Site | Path | $E_m^f$ | $E_m^r$ |
|---|---|---|---|---|---|
| Ni$_{0.5}$Co$_{0.5}$ | Ni | (6,6)–(4,8) | [100]NiNi–[010]NiNi | 0.27 | 0.21 |
| | | (6,6)–(8,4) | [100]NiNi–[010]CoNi | 0.16 | 0.20 |
| | | (6,6)–(6,6) | [100]CoNi–[010]CoNi | 0.17 | 0.28 |
| | Co | (6,6)–(6,6) | [100]CoCo–[010]CoCo | 0.27 | 0.38 |
| | | (6,6)–(8,4) | [100]NiCo–[010]CoCo | 0.17 | 0.21 |
| | | (4,8)–(4,8) | [100]NiCo–[010]NiCo | 0.21 | 0.27 |
| Ni$_{0.5}$Fe$_{0.5}$ | Ni | (6,6)–(6,6) | [100]NiNi–[010]NiNi | 0.42 | 0.37 |
| | | (5,7)–(6,6) | [100]NiNi–[010]FeNi | 0.74 | 0.37 |
| | | (6,6)–(6,6) | [100]FeNi–[010]FeNi | 0.38 | 0.20 |
| | Fe | (6,6)–(6,6) | [100]FeFe–[010]FeFe | 0.64 | 0.60 |
| | | (8,4)–(6,6) | [100]NiFe–[010]FeFe | 0.92 | 0.68 |
| | | (6,6)–(8,4) | [100]NiFe–[010]NiFe | 0.44 | 0.30 |
| Ni$_{0.8}$Fe$_{0.2}$ | Ni | (9,3)–(9,3) | [100]NiNi–[010]NiNi | 0.29 | 0.29 |
| | | (8,4)–(10,2) | [100]NiNi–[010]FeNi | 0.36 | 0.28 |
| | | (8,4)–(8,4) | [100]FeNi–[010]FeNi | 0.22 | 0.15 |
| | Fe | (10,2)–(6,6) | [100]FeFe–[010]FeFe | 0.74 | 0.41 |
| | | (10,2)–(8,4) | [100]FeFe–[010]NiFe | 0.34 | 0.45 |
| | | (6,6)–(10,2) | [100]NiFe–[010]NiFe | 0.26 | 0.46 |
| Ni$_{0.8}$Cr$_{0.2}$ | Ni | (9,3)–(8,4) | [100]NiNi–[010]NiNi | 0.22 | 0.28 |
| | | (8,4)–(8,4) | [100]NiNi–[010]CrNi | 0.10 | 0.26 |
| | | (9,3)–(8,4) | [100]CrNi–[010]CrNi | 0.00 | 0.50 |
| | Cr | (9,3)–(8,4) | [100]CrCr–[010]CrCr | 0.02 | 0.20 |
| | | (8,4)–(8,4) | [100]CrCr–[010]NiCr | 0.06 | 0.11 |
| | | (9,3)–(8,4) | [100]NiCr–[010]NiCr | 0.22 | 0.28 |

It is found that the migration energy of interstitials is smaller than that of vacancies. This is consistent with most metals, in which interstitials are highly mobile.[62] Compared with pure metals or conventional alloys with some solute atoms, the diffusion behavior in solid-solution alloys is significantly affected by the disorder. The most important factor is that the random atomic displacement in solid-solution along with the random relaxation induced by interstitial defects greatly changes the energy landscape for interstitial migration. Even for the simplest diffusion path investigated, the energies of reaction images are strongly dependent on the local environment. As a result, the diffusion barrier is very sensitive to the initial and final configurations, as well as the locations of the intermediate images.

Generally, the diffusion barrier for Ni and Co is very similar in Ni$_{0.5}$Co$_{0.5}$. The barriers determined in Ni$_{0.5}$Fe$_{0.5}$ and Ni$_{0.8}$Fe$_{0.2}$ are smaller for Ni interstitials than for Fe interstitials. Compared with the migration energy of 0.11 eV in pure Ni, we can see that most of the migration barriers are greatly elevated. These results therefore indicate that interstitial diffusion is much slower in the concentrated alloys than in pure metal. This conclusion is in agreement with previous MD results.[63] This effect is more prominent for Ni$_{0.5}$Fe$_{0.5}$. Depending on the initial and final locations of the interstitials, the diffusion barrier is not always located at the middle of the diffusion path but depends on the specific relaxation induced by the interstitials. In some cases, a complicated energy profile is observed with more than one energy extremum in the diffusion path. This mechanism is peculiar to concentrated solid-solutions with totally random atomic displacements.

The diffusion barrier for both Ni and Cr interstitials is relatively small in $Ni_{0.8}Cr_{0.2}$. In particular, there is a case that the Ni interstitials in a [100] CrNi dumbbell can migrate without any energy barrier to a [010] CrNi dumbbell at the nearest site, which has a 0.50 eV lower formation energy. These migration energies indicate that both Ni and Cr interstitials are highly mobile, and there is no preference. Compared with $Ni_{0.5}Fe_{0.5}$ and $Ni_{0.8}Fe_{0.2}$, these barriers are very small, indicating faster migration of interstitials in $Ni_{0.8}Cr_{0.2}$.

### *3.2.4 Comparison with EAM potentials for Ni-Fe alloy*

Three EAM potentials for Ni-Fe interactions developed by Bonny *et al.*[32-34, 64] are examined with regard to the description of defect energetics in Ni-Fe solid-solution alloys. To distinguish different versions, we denote the corresponding potentials by the year they were developed, that is Bonny2009,[32] Bonny2011,[33] and Bonny2013.[34] For each potential, the defect properties in $Ni_{0.5}Fe_{0.5}$ and $Ni_{0.8}Fe_{0.2}$ alloys are calculated to compare with *ab initio* results so as to evaluate the performance of these EAM potentials.

The formation energies of vacancies in $Ni_{0.5}Fe_{0.5}$ and $Ni_{0.8}Fe_{0.2}$ calculated using these three potentials are shown in Fig. 2 and compared with the corresponding *ab initio* results. Although we note that the results are obtained from a relatively small supercell, they present the main features of these three potentials. A detailed comparison is made by constructing a bigger 8×8×8 supercell with 2048 randomly distributed Ni and Fe atoms, from which each atom is removed to calculate the vacancy formation energy. The distribution of the calculated data overlaps exactly with the results shown in Fig. 2. This fact suggests that the 108-atoms SQS supercell represents the random structure of the alloy well.

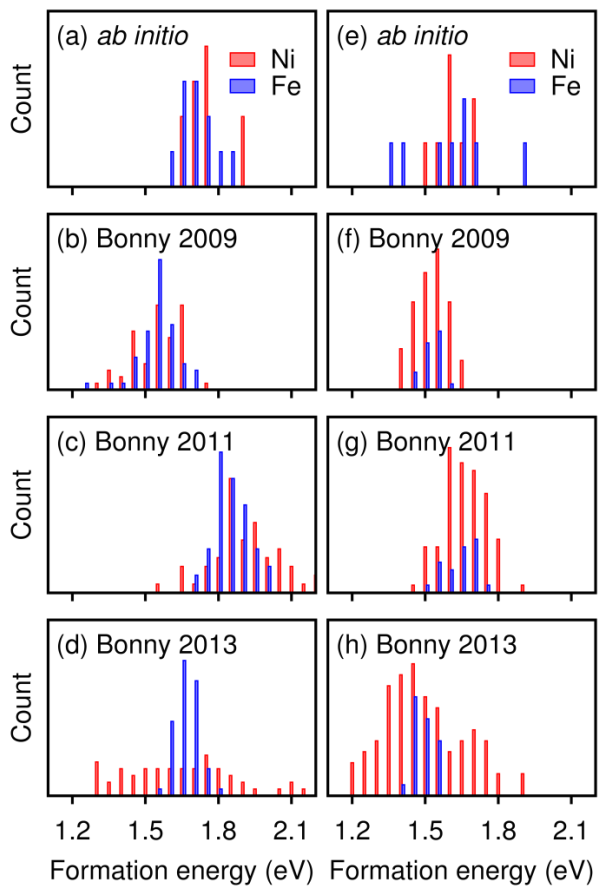

Fig. 2 (Color online) Formation energy of vacancy for $Ni_{0.5}Fe_{0.5}$ (left column) and $Ni_{0.8}Fe_{0.2}$ (right column) in a 108-atom SQS supercell calculated from *ab initio* calculations (a and e) and three EAM potentials: Bonny2009 (b and f), Bonny2011 (c and g) and Bonny2013 (d and h).

The formation energies for Ni and Fe vacancies in $Ni_{0.5}Fe_{0.5}$ significantly overlap both from *ab initio* calculations and EAM potentials. However, the distributions of formation energies from these three potentials exhibit distinct features. Both Bonny2011 and Bonny2013 potentials result in a wide-spread distribution of the formation energy of Ni vacancies, with the formation energy of Fe vacancies embedded in it. The highly overlapped distribution will have a profound influence on defect evolution when the potentials are used in MD simulations. In this case, the diffusion of vacancies of different types is greatly facilitated by their similar formation energies. For $Ni_{0.8}Fe_{0.2}$, the *ab initio* formation energies of Fe vacancies are spread over a large energy range, while the EAM potentials yield a narrower distribution. The results from *ab initio* calculations indicate the formation energies of vacancies in a Ni-Fe solid-solution are strongly composition dependent. However, it is hard to capture this feature with EAM potentials.

For [100] dumbbell interstitials, the configurations optimized by the *ab initio* method and by these three EAM potentials are different in solid-solution alloys. For example, while the Ni-Ni dumbbell is almost parallel to the *x* axis in *ab initio* calculations, that is not the result obtained by the EAM potentials as shown in Fig. 3. The optimized structure has an apparent intersection angle with the *x* axis. Consequently, the oblique dumbbell leads to asymmetric distortion of its nearest neighbors. In general, we found that the distortion of the [100] dumbbell from the Bonny2011 potential is less pronounced than those from Bonny2009 and Bonny2013.

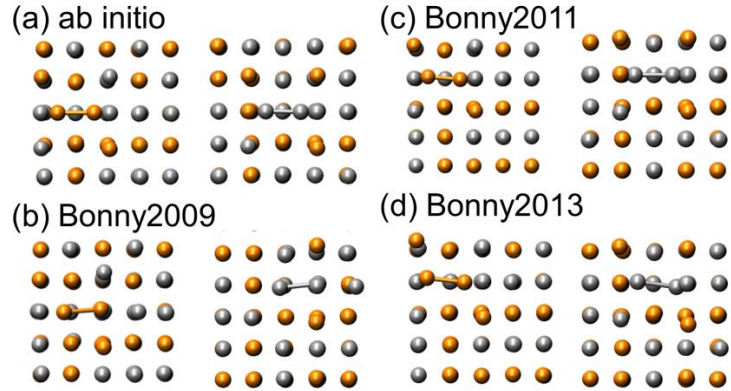

Fig. 3 (Color online) Configurations of [100] Ni-Ni (left column) and Fe-Fe (right column) dumbbell in $Ni_{0.5}Fe_{0.5}$ optimized from *ab initio* calculations and three EAM potentials within the same supercell.

The distribution of formation energies for the [100] dumbbell in the $Ni_{0.5}Fe_{0.5}$ and $Ni_{0.8}Fe_{0.2}$ alloys is presented in Fig. 4. The results from the EAM calculations are determined from a 2048-atom supercell to obtain a full sampling of the local environments of dumbbells. The dumbbell is introduced at each site, and the dumbbell structure is picked out after relaxation to calculate the formation energies. We have compared these results to those from a 108-atom supercell and the comparison confirmed that the formation energy of interstitials calculated from a small supercell is overestimated as expected due to the interaction between interstitial and its images. Nevertheless, the features of the distribution are basically the same but with a poor statistics in a 108-atom supercell. Since we would like to capture the true characteristics of these potentials, we chose the results from a 2048-atom supercell with good statistics to present the results.

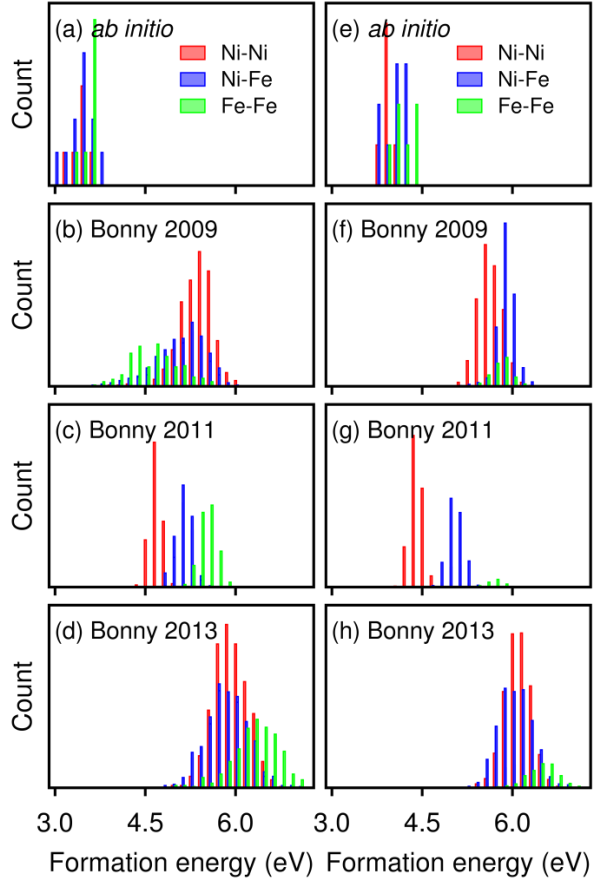

Fig. 4 (Color online) Formation energy of a [100] dumbbell in $Ni_{0.5}Fe_{0.5}$ (left column) and $Ni_{0.8}Fe_{0.2}$ (right column) determined from *ab initio* calculations (a and e) and three EAM potentials: Bonny2009 (b and f), Bonny2011 (c and g) and Bonny2013 (d and h). There are 26 data points in the *ab initio* results. For calculations based on EAM potentials, the dumbbell is introduced at each lattice site and then the dumbbell structure is picked out after relaxation to calculate the formation energies.

Figure 4 shows that the formation energies from the *ab initio* calculations are relatively small in both $Ni_{0.5}Fe_{0.5}$ and $Ni_{0.8}Fe_{0.2}$. However, all three EAM potentials predict larger formation energies with different distributions. The formation energy of the Fe-Fe dumbbell is the largest for Bonny2013 and Bonny2011. Moreover, Bonny2011 results in three separate distributions for the Ni-Ni, Ni-Fe, and Fe-Fe dumbbells, whereas Bonny2009 and Bonny2013 give largely mixed formation energies. In $Ni_{0.5}Fe_{0.5}$, while the formation energies for Fe-Fe dumbbells are smaller from Bonny2009 as compared to the other Ni-Ni and Ni-Fe dumbbells, an opposite trend is observed from Bonny2013. Compared with *ab initio* results, Bonny2013 generates a similar distribution for all dumbbells, although their energies are highly overestimated. Note that the small supercell (108 atoms) used in the *ab initio* calculations tends to overestimate the formation energies of interstitials. Thus the smaller formation energies of interstitials in $Ni_{0.5}Fe_{0.5}$ are not affected by the choice of supercells.

The diffusion barrier for vacancies in pure Ni is 1.17 and 1.11 eV calculated from Bonny2011 and Bonny2013, respectively, in line with the original report of 1.17[33] and 1.09 eV.[34] The value obtained from Bonny2009 is 0.98 eV, relatively smaller than that from Bonny2011 and Bonny2013. Note that the *ab initio* value is 1.01 eV, as indicated in Table I. Using an 108-atom supercell identical to that used in the *ab initio* calculations, we have calculated the diffusion barriers for both Ni and Fe vacancies using the CI-NEB method when they migrate toward all their nearest neighbor lattice positions. The calculated migration energies in $Ni_{0.5}Fe_{0.5}$ and $Ni_{0.8}Fe_{0.2}$ with these three potentials as well as the *ab initio* results are summarized in Fig. 5.

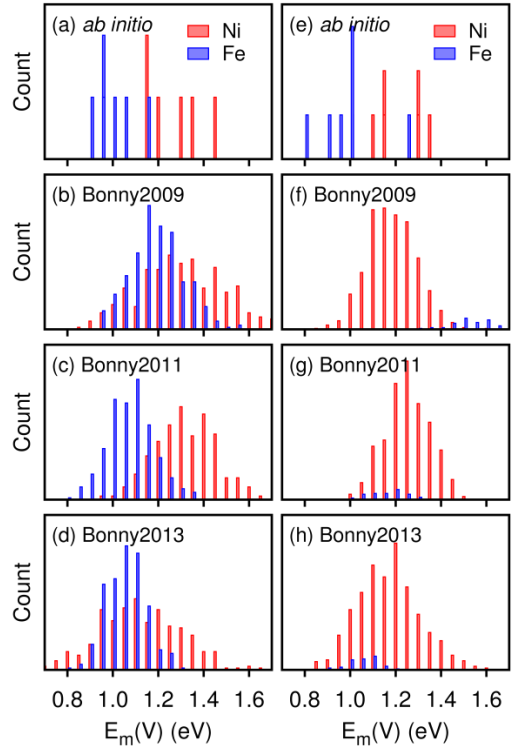

Fig. 5 (Color online) Migration energies of Ni and Fe vacancies in $Ni_{0.5}Fe_{0.5}$ (left column) and $Ni_{0.8}Fe_{0.2}$ (right column) determined from *ab initio* calculations (a and e) and three EAM potentials: Bonny2009 (b and f), Bonny2011 (c and g) and Bonny2013 (d and h).

As can be seen from Fig. 5, the migration energy of Ni vacancies is larger than that of Fe vacancies based on the *ab initio* calculations, which is also the feature from the Bonny2011 potential. In this case, the migration energies are somewhat separate for Ni and Fe vacancies. However, the distribution is different for Bonny2009 and Bonny2013. Both potentials predict that the migration energies of Fe vacancies greatly overlap those of Ni vacancies. The lower migration energy for Fe vacancies indicates that it is easier for them to diffuse in $Ni_{0.5}Fe_{0.5}$ and $Ni_{0.8}Fe_{0.2}$ than Ni vacancies. Nevertheless, Bonny2009 results in a larger migration energy for Fe vacancies in $Ni_{0.8}Fe_{0.2}$, contrary to other Bonny potentials. The higher barrier in $Ni_{0.8}Fe_{0.2}$ (Fig. 5f) can be attributed to the high migration energy for Fe in pure Ni which is 1.55 eV[32] from this potential. Here we can see that the energetics is strongly related to the alloy composition.

The diffusion of [100] dumbbell interstitials to the nearest lattice sites which result in the formation of another [010] dumbbell is investigated in a 2048-atom supercell using EAM potentials with the CI-NEB method. As a reference, the diffusion barrier in pure Ni is first calculated. The result is 0.13 eV from Bonny2009, 0.33 eV from Bonny2011, and 0.17 eV from Bonny2013. We see that the barrier from Bonny2011 is higher compared with the other two potentials. The diffusion barrier in $Ni_{0.5}Fe_{0.5}$ and $Ni_{0.8}Fe_{0.2}$ is calculated following the same method. To avoid distorted dumbbells that would give rise to a more complicated energy landscape, we first refine the dumbbell configurations with the least distortion in a 2048-atom supercell. Those dumbbells with displacements of less than 0.3 Å with respect to the perfect dumbbell structure obtained from *ab initio* calculations are used for the CI-NEB calculations. The diffusion barrier in $Ni_{0.5}Fe_{0.5}$ and $Ni_{0.8}Fe_{0.2}$ is shown in Fig. 6.

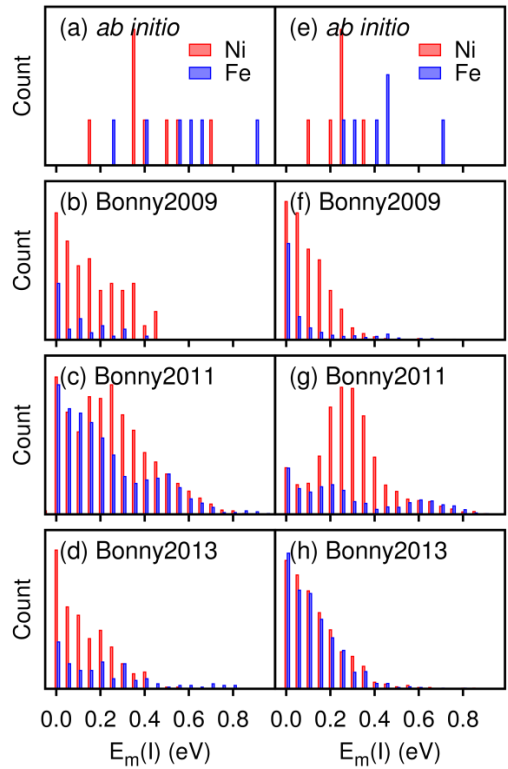

Fig. 6 (Color online) Migration energy of Ni and Fe interstitials in $Ni_{0.5}Fe_{0.5}$ (left column) and $Ni_{0.8}Fe_{0.2}$ (right column) determined from *ab initio* calculations (a and e) and three EAM potentials: Bonny2009 (b and f), Bonny2011 (c and g) and Bonny2013 (d and h).

The distortion of the dumbbell configurations from the Bonny2011 potential is the least among these three potentials. Therefore, more migration paths are established in the calculations. Compared with the migration energy of 0.11 eV in pure Ni, *ab initio* calculations predict higher barriers for interstitial migration in $Ni_{0.5}Fe_{0.5}$. However, the energies from these three potentials display different features. The barrier from Bonny2009 is clustered in the low energy-end from 0.00 to 0.50 eV. On the other hand, both Bonny2011 and Bonny2013 result in a broad distribution of migration energies. Note that the peak of the barriers in $Ni_{0.8}Fe_{0.2}$ from Bonny2011 (Fig. 6g) is related to the higher diffusion barrier in pure Ni. A considerable portion of the barrier predicted from Bonny2011 extends into the high-energy end. In comparison with the *ab initio* calculation, this reasonable agreement, especially at the high-energy end, suggests that Bonny2011 is more appropriate for simulation of interstitial diffusion in a Ni-Fe alloy.

Compared with pure Ni, in which the migration energy of vacancies and interstitials is a specific value, the migration energy in solid-solution alloys has a large spread. The results presented in Figs. 5 and 6 demonstrate that the distribution of migration energies for vacancies and interstitials in $Ni_{0.5}Fe_{0.5}$ has an overlap region. This will have a profound influence on the defect evolution in this alloy. When it is subjected to energetic ion irradiation, many Frenkel pair defects occur. The similar migration energies of vacancies and interstitials suggest their mobility is similar. This will greatly enhance the recombination of vacancies and interstitials and contribute to defect annihilation, making the alloy more irradiation-resistant.

# 4. Discussion

The distribution of defect energetics such formation energy and migration energy to a large extent determines the atomic diffusion behaviors under irradiation or thermal conditions. Taking $Ni_{0.5}Fe_{0.5}$ an example, the formation energies of both Ni and Fe interstitials are lower than those in pure Ni, which suggests strong binding of interstitial with lattice atoms. This effect, in combination with the high migration barriers calculated here, contribute to the sluggish diffusion of interstitials. In addition, the formation energies of Ni-Ni dumbbells are lower than Fe-Fe dumbbells and most migration barriers of Ni are larger than those of Fe, which indicate that the interstitials tend to diffuse through Ni sublattice. Consequently, the segregation of Ni is expected in $Ni_{0.5}Fe_{0.5}$ under non-equilibrium conditions. Comparing the results of $Ni_{0.5}Co_{0.5}$ and $Ni_{0.5}Fe_{0.5}$, it is found that most migration barriers in $Ni_{0.5}Co_{0.5}$ is lower than those in $Ni_{0.5}Fe_{0.5}$, suggesting that the defect diffusion coefficients in $Ni_{0.5}Co_{0.5}$ must be higher those that in $Ni_{0.5}Fe_{0.5}$. These barriers are all higher than those in pure Ni. Therefore, the diffusions of interstitial loops in these alloys are different with the order Ni>$Ni_{0.5}Co_{0.5}$>$Ni_{0.5}Fe_{0.5}$. This conclusion is in agreement with experimental observations.[65]

The defect energetics presented in this work is obtained at 0K. To include the temperature effect, the Gibbs formation energy should be calculated as $G_f = G_{defect} - G_{per} \pm \mu_D$, where $G_{defect}$ and $G_{per}$ is the free energy of defect-contained and defect-free crystals, respectively. $\mu_D$ is the chemical potential of defect D. The free energy should be determined by $G = E_0 + G_{elec} + G_{vib} + pV$, where $E_0$ is the ground state energy, $G_{elec}=E_{elec}-TS_{elec}$ is the electronic free energy, $G_{vib}$ is the vibrational free energy and *pV* is the product of pressure and volume. For defect formation, the change in volume is negligible and the *pV* term can be ignored. The electronic free energy can be calculated from electronic density of states. The vibrational free energy is determined by the phonon density of states, which requires accurate phonon calculations. For disordered system, the determination of both electronic and vibrational free energy is very difficult and time consuming. Thus we use the formation energy at 0K to analyze the defect properties. Nevertheless, the results are still instructive to analyze the defect behaviors in these alloys. Specifically, the determined alloy structure details can be used to predict the defect properties. For example, the first neighbor distances in $Ni_{0.5}Fe_{0.5}$ are the largest for Fe-Fe, which suggests that Fe has a large size than Ni in $Ni_{0.5}Fe_{0.5}$. This explains why Ni interstitials are energy preferred in this alloy. Although the distance will change along with increasing temperature due to thermal expansion, the conclusion that Ni is smaller than Fe in $Ni_{0.5}Fe_{0.5}$ is still valid. Thus these results still indicate that a preferable diffusion of Ni than Fe.

The evaluation of available EAM potentials for Ni-Fe regarding to point defect energies suggests that although these EAM potentials give relatively consistent results about the properties of vacancies, the formation energies for interstitials are overestimated. Moreover, these potentials predict distorted dumbbell interstitials that are different from those determined from *ab initio* calculations. Since these energies are directly related to the defect behaviors such as defect diffusion and agglomeration, the analysis of MD results based on EAM potentials should be examined carefully.

# 5. Conclusion

The properties of point defects including formation energies and migration energies in Ni-based solid-solution alloys including $Ni_{0.5}Co_{0.5}$, $Ni_{0.5}Fe_{0.5}$, $Ni_{0.8}Fe_{0.2}$ and $Ni_{0.8}Cr_{0.2}$ are studied using first-principles calculations based on DFT. Atomistic simulations using EAM potentials are also carried out for Ni-Fe alloys to evaluate their predictive capability regarding defect properties. The chemical disorder is taken into account by the SQS method. An efficient and accurate method of calculating the elemental chemical potentials in random alloys is proposed based on the distribution of nearest neighbors around various atomic sites. The results show that most formation energies of vacancies in concentrated alloys are larger than those in pure Ni, while the formation energies of [100] dumbbell interstitials in $Ni_{0.5}Co_{0.5}$ and $Ni_{0.5}Fe_{0.5}$ are much smaller than those in Ni. These properties are shown to be closely related to the details of alloy structures. For migration barriers, it is found that $Ni_{0.5}Fe_{0.5}$ and $Ni_{0.8}Fe_{0.2}$ have the largest effect in modulating the defect migration energies. In addition, the migration energies of vacancies and interstitials in $Ni_{0.5}Fe_{0.5}$ have a region of overlap, an indication of enhanced defect annihilation in $Ni_{0.5}Fe_{0.5}$.

# Acknowledgement

This work was supported as part of the Energy Dissipation to Defect Evolution (EDDE), an Energy Frontier Research Center funded by the U.S. Department of Energy, Office of Science, Basic Energy Sciences. The authors would like to thank German D. Samolyuk and Laurent K. Beland for their insightful comments to this work.